# Direct Measurement of the Magnitude of van der Waals interaction of Single and Multilayer Graphene

Yu-Cheng Chiou[1*], Tuza Adeyemi Olukan[2*], Mariam Ali Almahri[2], Harry Apostoleris[2], Cheng Hsiang Chiu[2], Chia-Yun Lai[2,], Jin-You Lu[2], Sergio Santos[3], Ibraheem Almansouri[2], Matteo Chiesa,[2,4]

[1] Topco Scientific Co. Ltd., No.483, Sec. 2, Tiding Blvd., Neihu, Taipei City 11493, Taiwan

[2]Laboratory for Energy and NanoScience (LENS), Khalifa University of Science and Technology, Masdar Institute Campus, Abu Dhabi, UAE

[3]Future Synthesis AS Uniongata 18, 3732 Skien, Norway

[4]Arctic Renewable Energy Center (ARC), Department of Physics and Technology, UiT, Norway

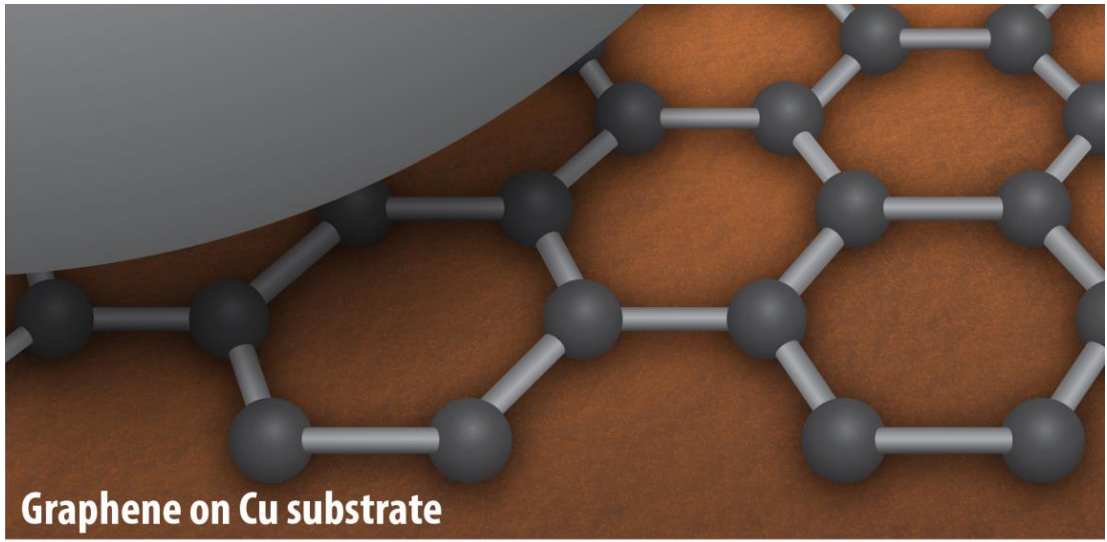

Graphene on Cu substrate

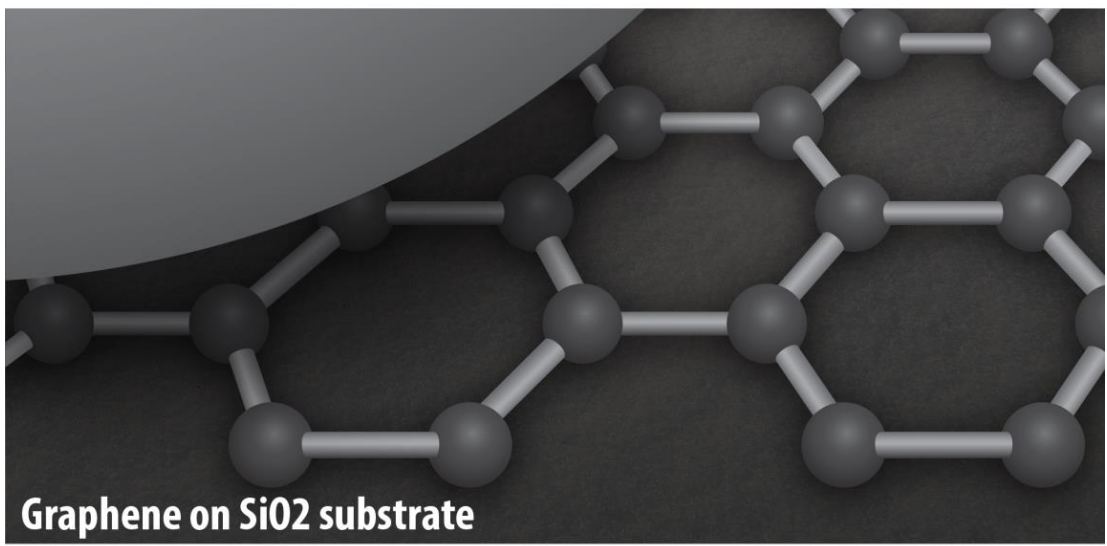

Graphene on SiO2 substrate

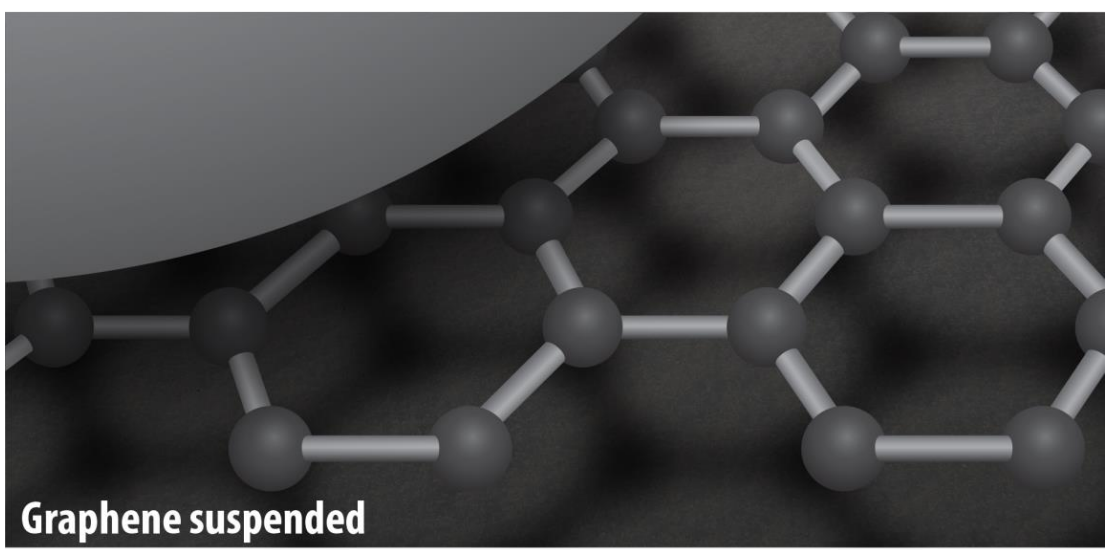

Graphene suspended

**Abstract**

Vertical stacking of monolayers via van der Waals assembly is an emerging field that opens promising routes toward engineering physical properties of two-dimensional (2D) materials. Industrial exploitation of these engineering heterostructures as robust functional materials still requires bounding their measured properties so to enhance theoretical tractability and assist in experimental designs. Specifically, the short-range attractive van der Waals forces are responsible for the adhesion of chemically inert components and are recognized to play a dominant role in the functionality of these structures. Here we reliably quantify the the strength of van der Waals forces in terms of an effective Hamaker parameter for CVD-grown graphene and show how it scales by a factor of two or three from single to multiple layers on standard supporting surfaces such as copper or silicon oxide. Furthermore, direct measurements on freestanding graphene provide the means to discern the interplay between the van der Waals potential of graphene and its supporting substrate. Our results demonstrated that the underlying substrates could enhance or reduce the van der Waals force of graphene surfaces, and its consequences are explained in terms of a Lifshitz theory-based analytical model.

## Introduction

The development of graphene and the entire class of 2D materials[1] over the last decade has raised tantalizing application possibilities that leverage on the unique physics of 2D crystal structures to engineer materials at the nanoscale. In recent years interest in this area has turned towards the concept of "van der Waals heterostructures[2]," in which multiple 2D layers are stacked with precise orientations to yield the desired properties, such as for example electronic band gaps which can be tuned by varying the constituent layers and their orientation[3] 2D layers can be seen as building blocks from which novel atomic scale metamaterials are constructed[4], opening up a new paradigm of "2D manufacturing" whereby new structures are engineered from the atomic level up by the combination of these 2D building blocks to yield the desired properties. These structures are unique in that the component layers join by other than chemical bonds. The socalled "glue" that binds these blocks is the ubiquitous van der Waals VdW forces[5] that arise from plane to plane interactions. It is therefore bring 2D manufacturing of materials to fruition. In this work we quantify the VdW interaction of CVD-grown graphene by means of the Hamaker coefficient, a parameter that summarizes the strength of such interactions. Hamaker[6] demonstrated that the van der Waals force strength between two bodies could be split into a purely geometrical component and a coefficient that depends solely on material chemistry i.e. polarizabilities and number densities of the atoms in the two interacting bodies[6]. This factor has become known as the Hamaker parameter, that is here treated as a constant because of the small interaction range that we consider, and denoted as A. Lifshitz[7] presented a more rigorous approach that incorporated the many-body effects neglected in Hamaker's approach and which is based on a thermodynamic consideration of the interacting bodies as a continuum described by their dielectric properties. As the van der Waals interaction ultimately results from the fluctuations of the electromagnetic field between two macroscopic bodies, the Hamaker coefficient as described by Lifshitz serves as a summary parameter quantifying the strength of this interaction for a given material system[8]. As such it offers a more general picture of the surface characteristics than the measurement of a particular physical quantity such as adhesion or surface energy.

Despite in the clear relevance of these forces to understand of the interactions of 2D structures, the Hamaker constant -and indeed the VdW force profiles that it generates -of graphene and other 2D materials remains poorly studied. In the present work we directly quantify the van der Waals interactions of graphene surfaces by using the observables of a newly developed bimodal AFM[9] methodology to map the Hamaker coefficient in the non-retarded approximation regime[8]. We note that an important factor in surface characterization of 2D materials, which we account for in this study, is the impact of the substrate on the measured values. As the sample thickness is on the order of Angstroms, surface force measurements may be influenced by the underlying substrate as well as by the sample. We perform measurements of samples on a variety of substrates, to evaluate the impact of substrate on the measured VdW strength of the graphene surfaces. We note that the experimental methodology followed to produce the samples might be of relevance for future experimentation and results critical in isolating to real measured forces.We have employed nanofabrication techniques to create patterned substrates that support regions of free-standing graphene where the graphene-substrate distance is on the order of microns. By performing measurements in these suspended regions, we characterize the graphene itself, removing the effect of the substrate. With these results we have succeeded in directly measuring the VdW strength of graphene surfaces on the nanoscale .

## Experiment

The graphene for our measurements is grown on Cu substrates in-house via chemical vapor deposition (CVD). Varying the gas precursor flow rates as described in the methods section controlled the number of layers. Confirmation of the number of graphene layers was done by Raman spectroscopy via the ratio of the 2D and G peaks in a Raman spectrum, which varies from about 3 in single-layer graphene and decreases to less than 1 in multilayer samples. We take Raman spectra of both graphene-on-Cu (as-grown) and graphene transferred onto both flat $SiO_2$ and patterned $SiO_2$ substrates that support regions of suspended graphene as described in the introduction. The patterned substrate was created via focused ion beam etching to create a pattern of "holes" over which the graphene layers are transferred (details in supplementary). Atomic force microscopy (AFM) can then measure the Hamaker of graphene alone in these suspended regions, without influence of the substrate.

After Raman measurement, as-deposited and transferred graphene samples were put into the AFM where $100\text{x}100\,\text{nm}^2$ maps of the surface were collected in bimodal operation (see Methods section). The mapped regions fell inside the area where the Raman spectrum was taken. The Hamaker coefficient was mapped based on a method described in previous work and summarized in the methods section. In summary, the Hamaker can be derived from raw bimodal AFM observables via the relationship to direct observables as:

$$A = -\frac{3\pi k_2 A_{02} \cos(\varphi_2)}{0.83 R Q_2 A_2} \sqrt{d^5 A_1} \qquad (1)$$

where R is the tip radius, $k_m$ and $Q_m$ the spring constant and quality factor, respectively, $d_{min}$ (THE EQUATION HAS NO Dmin) the minimum distance of approach, $A_m$ the oscillation amplitude of the $m^{th}$ mode and $A_{0m}$ the free amplitude of the $m^{th}$ mode. In addition to the bimodal mapping, force spectra were taken in standard single-mode operation. The force vs. distance profiles were reconstructed using the Seder-Jarvis-Katan method from which an effective Hamaker can be obtained by fitting the attractive part of the force with an inverse squared power law. In addition to these experimental measurements we performed density functional theory (DFT) simulations to generate, from first principles, force-distance profiles for 1-, 2- and 3-layer suspended graphene. The simulations assisted in our interpretation by providing cause-effect controllable relationships.

**Results**

The value of the Hamaker constant for graphene on Cu is mapped as shown in Figure 1. The three Hamaker maps, Figure 1a, b, c, show regions of single, double and multi-layer graphene as confirmed by Raman spectra, Figure 1e. Hamaker values for each pixel are extracted and the distribution of these values for each of the three samples is reported in Figure 1d. The results show for the first time a clear difference between the VdW strength of mono, bi and multi-layer graphene. Moreover, the Hamaker maps provide an indication of graphene continuity at far higher resolution (nm scale) than Raman spectroscopy. This provides additional insight into the origin of the observed Hamaker values. For example, the regions in the maps of single and multi-layer graphene where the measured Hamaker abruptly changes, are likely to represent Cu grain boundaries that would impact graphene growth and the measured strength of the VdW interaction. Clearly the mean Hamaker values are affected by the substrate, as the variations in the measured values between the grain boundaries and bulk crystal regions demonstrates. Thus, the Hamaker values are to be looked upon as effective values that include the effect of the substrate. Considering further the influence of the substrate, we note that the substrate itself can be modified during the CVD process (e.g. by promoting hydrogenation of the surface). While it is clear that the influence of the substrate on measured graphene surface properties presents an additional challenge for characterization, it also provides an extra degree of freedom to selectively modifying the effective properties of graphene. That is, one must specify the substrate in order to understand the properties of graphene. The implication is that graphene, in that sense, should not be considered as the whole of the physical entity from which properties arise, and has to be reported as a substrate-graphene system instead. Furthermore, growth processes may also play a role, the differences in the measured surface properties between different graphene samples may, in this understanding, be in part related to differences in the substrate induced by the variations in the growth process. The

trends observed in Figure 1 are confirmed with different samples and different tips of radius R <
7 nm[10].

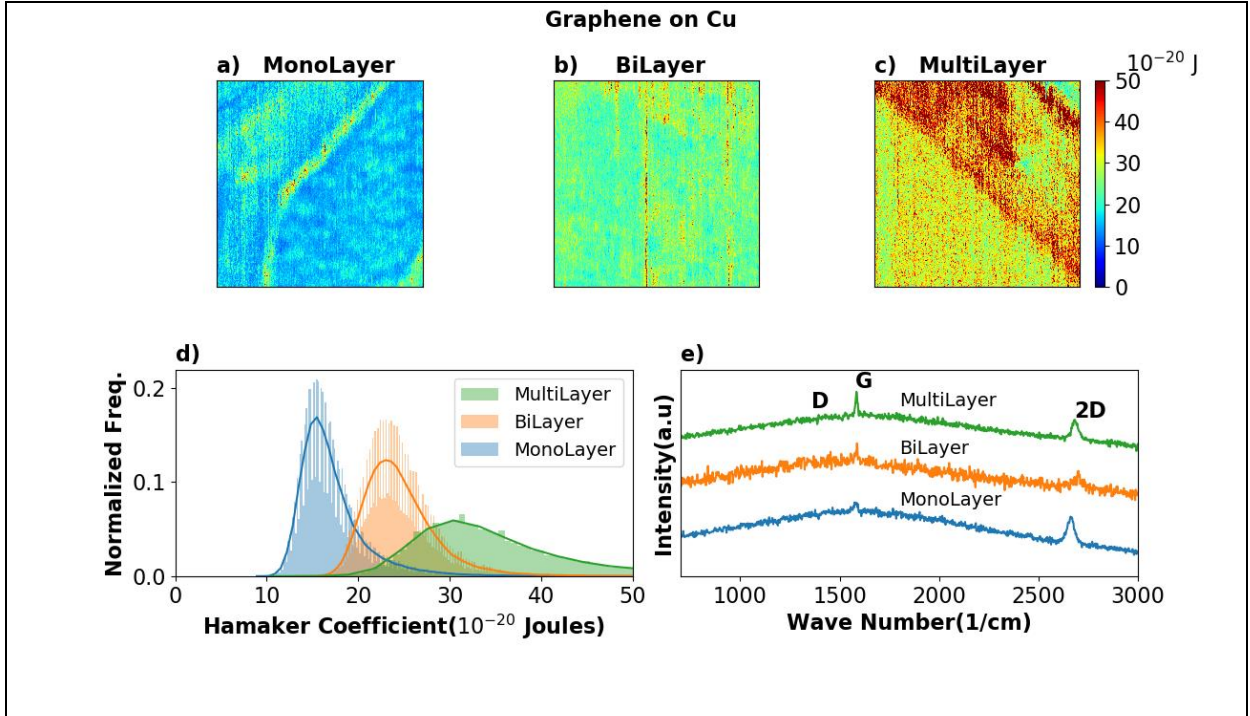

Figure 1 a, b, c) 100 nm × 100 nm Hamaker maps for monolayer, bilayer and multilayer graphene on Cu substrate respectively. d) Distribution of Hamaker values belonging to monolayer, bilayer and multilayer graphene. The Hamaker distributions present the raw data where no filter is applied see SI. e) Raman spectra plots of regions where monolayer, bilayer and multilayer graphene are identified. No filter is applied and the curves represent the raw data collected. The 532nm laser source has a spot size with radius ≈ 5 μm making the area extension several order of magnitude larger than the Hamaker maps see SI.

Figure 2 shows a summary of the Hamaker distributions and Raman spectra for mono, bi and multilayer samples on all of the substrates considered, namely Cu in Figure 2a, b, $SiO_2$ in Figure 2c, d and freestanding graphene Figure 2e, f (see supporting information). One of the first observations one can derive from the results in Figure 2 is the reduced strength of the VdW forces for non-metal substrate or freestanding graphene. Strikingly, the measured Hamaker for graphene on SiO$_2$ is lower than that for the suspended graphene with no substrate at all. We hypothesize that this reduced Hamaker might relate to variations in the dielectric/refractive properties of the effective Silicon-graphene surface alone, rather than in material density since it is clear that the presence of the substrate would lead to an additive atom density increase according to Hamaker's method. Roughly speaking wave interference might lead to absorption and emission resonance affecting the effective dielectric constant of the Hamaker-substrate surface. We will provide experimental evidence supporting this claim below. One can also notice in Figure 2d, a strong peak just below 1000cm$^{-1}$, which indicates the presence of the SiO$_2$ substrate. The strong Raman signal in the region 800-1000cm$^{-1}$ may also indicate the presence of PMMA residual. This last assumption cannot be entirely ruled out. On the other hand, the dominating contribution is from the SiO$_2$ substrate itself. As explained in the SI file, the transfer process is tedious and may also leave residuals of water between the graphene and the SiO$_2$ in addition to PMMA. The presence of a strong SiO$_2$ signal though might suggest that minimal water residuals are present at the graphene SiO$_2$ interface. The presence of residual could further provide some screening effect and thus reduce the measured effective Hamaker, however this reduction in the effective Hamaker on non-metal substrates can be explained even without invoking screening effect of the PMMA residue or water, as will be elaborated later in the manuscript

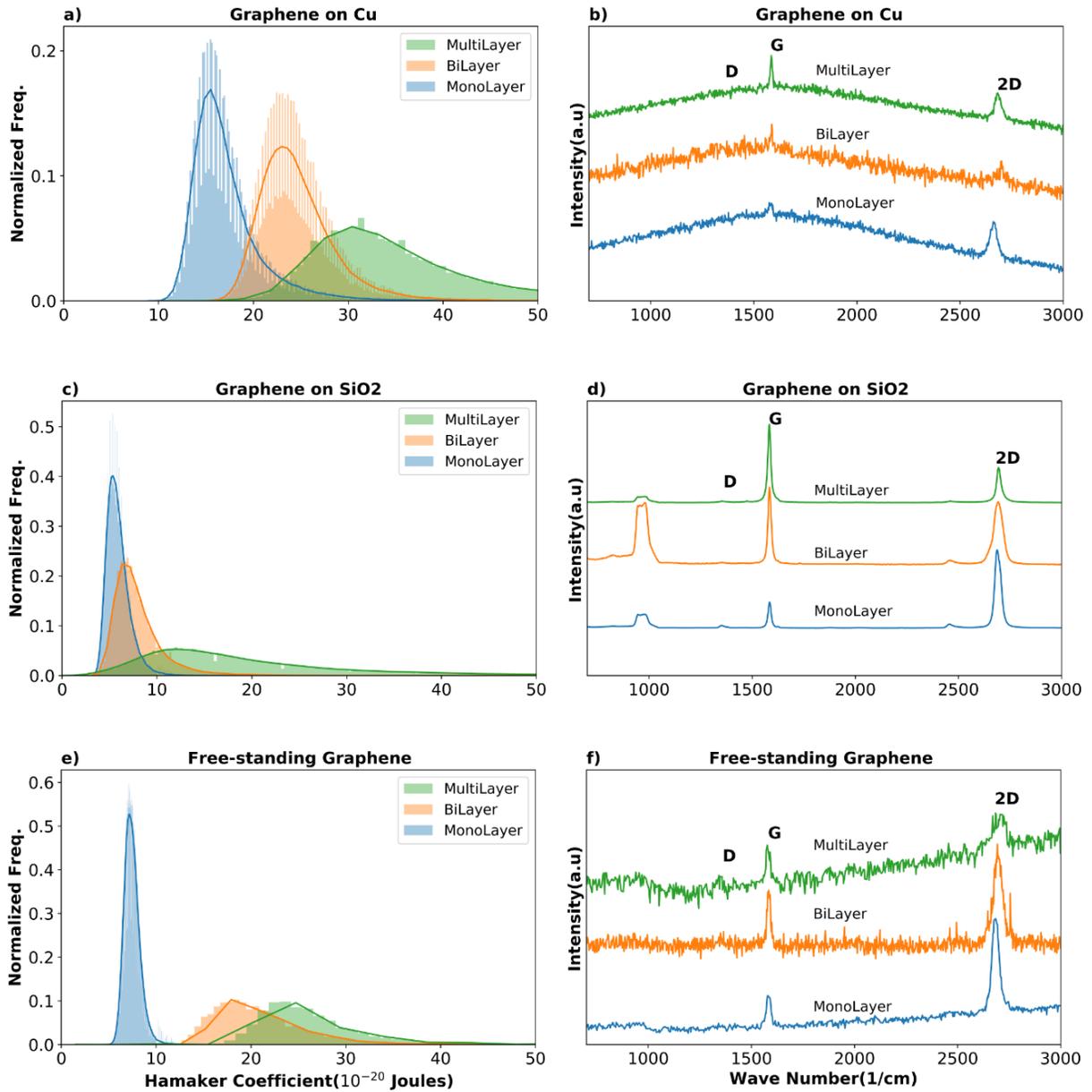

Figure 2 a, c, e) Distribution of Hamaker values belonging to the different Hamker maps directly measured by bimodal AFM and comparison to Raman spectra for graphene on Cu, SiO₂ and free standing b, d, f). It is worth noticing that the Raman spectra represent an area that is almost 3 orders of magnitude greater that the area observed in our AFM maps.

For SiO₂ substrate, Figure 2c, and the freestanding graphene Figure 2e, the distribution for the monolayer sample is very sharp indicating a high homogeneity of the graphene film, while in the case of bilayer and multilayer samples the distributions are much broader indicating possible heterogeneity in the graphene coverage. As before, this heterogeneity can be observed thanks to the high spatial resolution of our Hamaker maps where regions of single, double and multi-layer graphene can be observed see Figure 3a, b, c and SI for more details. The suspended graphene has no noticeable $SiO_2$ signal reinforcing the fact that the graphene is sufficiently isolated from the substrate and can be considered as free standing. Figure 2e also exhibits a clear distinction between monolayer and bilayer graphene directly quantifying the interlayer van der Waals interaction strength. The distinction between bilayer and multilayer graphene is less pronounced for the suspended graphene, but similar trends are observed in all the samples. Furthermore it is worth reflecting on how the substrate influences the Hamaker maps versus the results in the Raman spectrum. In the Raman, the Cu substrate presence is detected as noise that reduces the overall signal-to-noise ratio and cannot be effectively decoupled. However, in the Hamaker, the substrate increases, in the case of metal substrates, and decreases, in the case of non-metal substrate, the total measured value of the effective Hamaker coefficient, providing a means to quantify the strength of the probe-substrate interaction. Therefore this approach offers the possibility of separating the measured surface properties into a graphene-dependent and a substrate-dependent component.

Before proceeding in such an attempt our results are corroborated by collecting force-distance profiles for the suspended graphene (see Figure 3d). The Hamaker values derived by fitting the force-profile in Figure 3d with an inverse squared power law (in line with the assumption of a force distance relationship $F \approx RA/d^2$) were 11 e$^{-20}$ J, 17 e$^{-20}$ J and 36 e$^{-20}$ J for the monolayer, bilayer and multiple layer respectively. These values fall inside the Hamaker coefficient distributions reported in Figure 2e.

We further corroborate our experimental observations by means of DFT calculations. The DFT derived interaction energy and force-distance profiles are reported in the Figure 3d and SI respectively, where interaction energy $\Delta E(d) = E_{tot}(d) - E_{tip} - E_{graphene}$ between the tip and the graphene layers are directly taken from the energy difference between the total system with different tip-surface distance and individual tip/graphene. The force profiles are taken from the gradient of total energy of the tip-graphene system, which is given by $\boldsymbol{F} = -\boldsymbol{\nabla}(E_{tot}(d))$. As the number of graphene layers increases, the magnitude of the adhesion force and Hamaker coefficient also increase. This is in agreement . with our experimental observations.

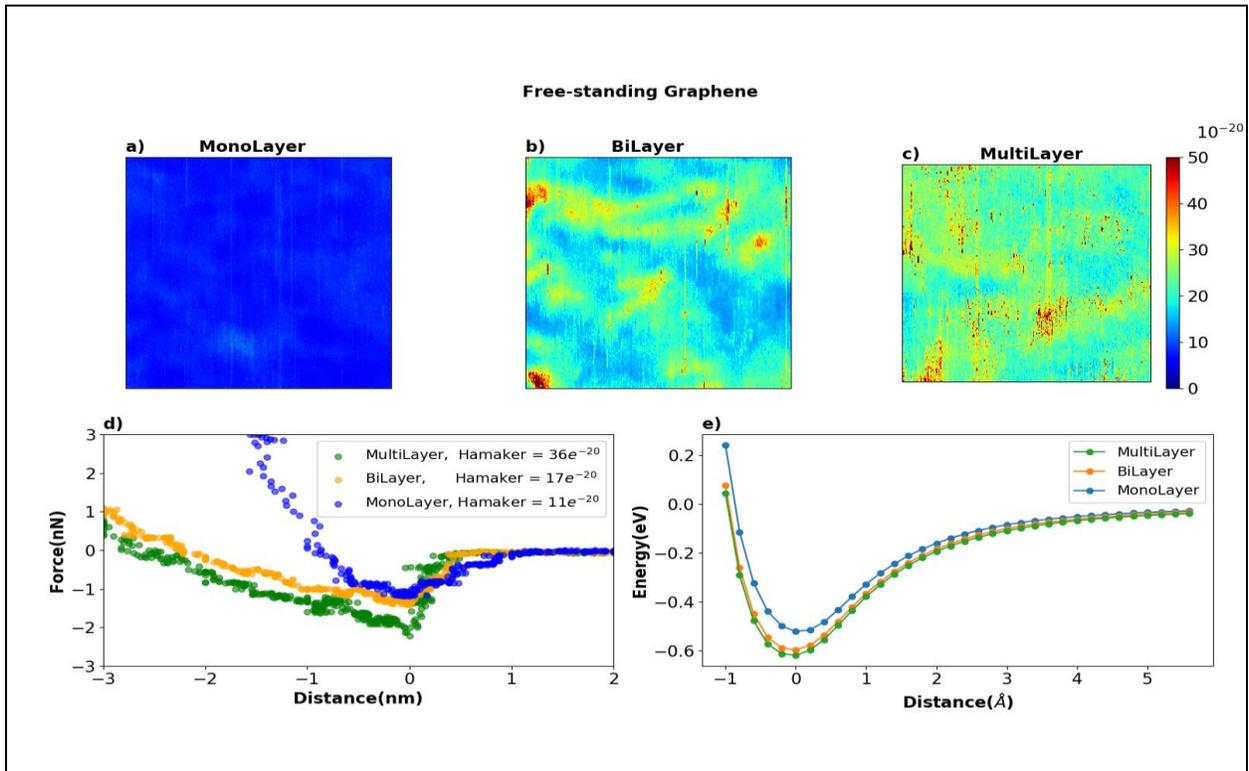

Figure 3 a, b, c) 100 nm × 100 nm Hamaker maps for free-standing monolayer, bilayer and multilayer graphene respectively. d)Force profile for Monolayer, Bilayer and Multilayer suspended graphene samples reconstructed by AM-AFM. e) DFT predicted interaction energy vs distance of tip to freestanding graphene with different layers see SI for more information.

While useful for confirming trends, the DFT simulations are not so well suited to investigate the role of the substrate; consider the computational intensiveness of the problem. In order to discern the effect of the substrate we revert to an analytical model employing two key approximations:

$$F = -\frac{A_{eff} R}{6L^2} \cong -\frac{A_{402} R}{6L^2} - \frac{A_{102} R}{6(L+b)^2} \qquad (2)$$

$$F = -\frac{A_{eff} R}{6L^2} \cong -\frac{A_{402} R}{6L^2} - \frac{A_{102} R}{6(L+b)^2} \left( \frac{n_1^2 - n_4^2}{n_1^2 + n_4^2} / \frac{n_1^2 - n_0^2}{n_1^2 + n_0^2} \right) \qquad (3)$$

where $A_{eff}$ is the effective Hamaker, $A_{402}$ is the Hamaker between AFM tip and graphene in air (obtained by our experiment see Figure 2 e since data is no available) and $A_{102}$ is the Hamaker between the $SiO_2$ tip and Substrate in air. Equation 2 and 3 refer respectively to metal and non-metal substrates (see SI for the derivations). Let R be the tip radius, $L$ be the distance between graphene and AFM tip, $b$ the graphene thickness and η the retractive index where the meaning of the suffixes 0-4 is given below. The values of $L$ and $b$ corresponding to monolayer graphene sample are 3.8Å and 3.35 Å, respectively, as predicted by our DFT simulations. In this work, materials 0, 1, 2, and 4 are vacuum/air, substrate (Cu or $SiO_2$), $SiO_2$, and graphene, respectively. To calculate the effective Hamaker constant, the $A_{402}$ is taken from the experimental values of the suspended graphene since to our knowledge no value is available in the literature (see Table 1). The values of $A_{102}$ are taken from the Hamaker constants of Cu and $SiO_2$, respectively. The comparison with the experimental AFM derived Hamaker constants, are shown in Table 1. Despite the great deal of assumptions necessary to feed equations (2) and (3), our calculations on copper agree surprisingly well with experimental data. On the $SiO_2$ substrate, the prediction qualitatively agrees with the experimental data despite being less accurate. This reduced accuracy may be due to the residuals left on the surface or between the surfaces by the transfer process (PMMA and water respectively), that we do not account for in our model. Nevertheless, the model captures the fact that the effective Hamaker constant of graphene on $SiO_2$ substrate is weaker than those of suspended graphene layers, something that can be explained by the negative value of $(\frac{n_1^2 - n_4^2}{n_1^2 + n_4^2} / \frac{n_1^2 - 1}{n_1^2 + 1})$ for $SiO_2$ substrate (refractive index $n_1 \sim 1.45$), which indicates that the VdW force on top of the graphene layer is reduced by the $SiO_2$ substrate while it is enhanced on the metal substrates.

Table 1. A comparison of effective Hamaker constants between experiment and our theoretical prediction.

| Method | Substrate | Monolayer $e^{-20}$ J | Bilayer $e^{-20}$ J | TriLayer $e^{-20}$ J |
|---|---|---|---|---|
| Experiment | Suspended | 7.2 | 17.99 | 24.7 |
| | Cu | 16 | 23.03 | 30 |
| | $SiO_2$ | 5.4 | 6.34 | 12 |
| Calculation | Cu | 15.22 | 21.71 | 26.87 |
| | $SiO_2$ | 4.16 | 16.58 | 23.92 |

This also explains in terms of van der Waals potentials the wetting transparency of monolayer graphene reported in the literature since the measured effective Hamaker for the monolayer graphene on the Cu substrate, reported in Table 1, is strikingly similar to the tabulated value for Hamaker of the $SiO_2$-Cu pair, i.e. 15.6.

Furthermore, our findings support a recently proposed hypothesis that tries to provide an analytic thermodynamic criterion for subsequent layer CVD-growth that depends among others on the vdW interaction energies. The large number of reports on CVD-growth of graphene on metal substrate shows how this process is routinely achieved and extendable to large-scale fabrication thanks to the strong van der Waals potential of the metal substrates support that constructively help in the growth of monolayer graphene. This is not the case for $SiO_2$ substrates. Equation 3 points out that if not properly matched (refractive index), the substrate may reduce the strength of the van der Waals interaction thereby inhibiting direct CVD-growth of single monolayers. Despite the oversimplification of equation 3 and realizing that there are many more effects that might independently participate in the overall phenomena, equation 3 can be used as a general rule of thumb to assess whether direct CVD-growth on non-metal substrate is feasible simply by looking at the substrate refractive index.

**Conclusions**

Direct measurement of the strength of the VdW forces of graphene has long been challenging, posing a problem for the application of graphene and other 2D materials, whose most promising potential uses (e.g. "van der Waals stack" applications) often depend on the precise manipulation of surface forces. In this study we have leveraged on the power of atomic force microscopy to measure the Hamaker coefficient of in-house-grown CVD graphene samples, thereby quantifying the strength of the VdW interaction. The use of AFM allows the Hamaker to be directly probed with nanoscale resolution, addressing a significant difficulty in prior studies which is the inability to distinguish "true" properties of single, double and triple layer graphene from the average properties that a measurement with low spatial resolution will detect when characterizing a sample with micro- or nano-scale variations in the graphene thickness. To resolve a further challenge, the effect of the substrate on the measured surface forces, we have conducted studies on different substrates, including a specially-fabricated substrate that supports regions of suspended graphene where the effect of the substrate is eliminated. DFT calculations are used to corroborate the measurements and show qualitative agreement with our observations on suspended graphene. An analytical model is developed from the theory of Lifshitz to explain the observed values and provide a means of quantifying the impact of the substrate, leading to the recognition that the substrate may, depending on its dielectric properties, either significantly reduce or enhance the VdW interaction measured at the graphene surface. The AFM-based techniques described here, as they can be easily implemented in any laboratory, without sophisticated equipment or involved sample preparation, provide a means of efficiently collecting large quantities of data that will be valuable in resolving the persistent questions and uncertainties regarding the surface properties of graphene and other 2D materials.

**Methods**

Graphene growth:

We use a planar TECH planarGROW-2S thermal CVD system with parallel heaters to synthesize graphene on Cu foil substrates. Sigma Aldrich Cu substrates (25 μm thick, 2× 2 cm2, 99.999% pure) for all experiments. Cu foils were cleaned prior to each growth by sonication for a total of 10 min in acetone, IPA, and DI water sequentially followed by drying (nitrogen blowing) before loaded into the CVD chamber. The growth of mono-layer graphene was carried out with a relatively low methane flow rate of 2 sccm and a hydrogen flow rate of 20 sccm at 1000 °C. The process starts by annealing the copper substrate surface for 15 min using hydrogen flow (5 sccm, 0.1 Torr, and 1000 °C). Following the high-temperature annealing, methane is introduced to the process for 120 minutes (2 sccm, 0.2 Torr, and 1000 °C), with a hydrogen flow of 20 sccm. Before cooling the chamber, an additional growth process was introduced to obtain graphene layers stacking in one sample. A higher methane flow (20 sccm, 2 Torr, and 1000 °C) was then introduced for 5 minutes while the hydrogen flow remained the same at 20 sccm. After the two-step growth process, the exposure to methane and hydrogen, the sample was cooled to room temperature and removed. See SI for more information.

Table 1: Graphene deposition parameters

| Step | Time (minutes) | Hydrogen | Methane | Temperature | Pressure |
|------|----------------|----------|---------|-------------|----------|
| 1st step | 120 | 20 | 2 | 1000 °C | 0.2 |
| 2nd step | 5 | 20 | 20 | 1000 °C | 2 |

Graphene transfer:

We use a standard process whereby the graphene-on-Cu sample is spin coated with PMMA and then immersed in an aqueous solution of ferric cloride to etch the Cu. When the Cu is dissolved in the etchant, the graphene/PMMA stack is transferred into DI water to remove the unwanted Cu residues. Finally the exposed graphene side is brought into contact with the $SiO_2$ substrate and the PMMA is etched away in acetone.

Raman Analysis:

A Witec Alpha 300 RAS Raman spectroscopy with 532nm laser source is employed in our experiments. During the measurement, the laser spot size diameter is kept constant and measures approximately 5 µm.

Bimodal operation:

An Asylum Research Cypher AFM and standard AC240TS cantilevers with spring constants k, Q factors and resonant frequencies f of $k_{(1)} \approx$ 1.5-3 N/m, $k_{(2)} \approx$ 60-90 N/m, $Q_{(1)} \approx$ 100, $Q_{(2)} \approx$ 400, $f_{(1)} \approx$ 70 kHz and $f_{(2)} \approx$ 450 kHz were employed. The subscripts stand for mode number, i.e. 1 and 2. Standard AC240TS cantilevers were oscillated at the first 2 modal resonance frequencies while the frequencies were determined with thermal analysis when the cantilevers were close to the sample surface (~30 nm). Cypher AFM was set to operate in attractive regime, that is, first mode free amplitude $A_{01}$ was set at ~$0.5A_c$ and the setpoint was set at ~$0.7A_{01}$[9b, 11]. First two modes' oscillation amplitude and phase channel ($A_1$, $A_2$, $\varphi_1$, and $\varphi_2$) were recorded. We then employed

$$d_{\min} - \left[ \left[ \frac{3\pi k_{(2)} A_{02} Q_{(1)} \cos\phi_{(2)}}{0.83 k_{(1)} A_{01} Q_{(2)} \cos\phi_{(2)}} \right]^{2/3} \frac{A_1}{(A_2)^{2/3}} \right] d_{\min}^{2/3} + 2A_1 = 0 \qquad (4)$$

and Eq. 1 to obtain Hamaker coefficient values. See SI for more information.

Force reconstruction:

A Cypher AFM from Asylum Research was operated in amplitude modulation (AM) mode and standard AC240TS cantilevers (k $\approx$ 2N/m, Q $\approx$ 100, and f0 $\approx$ 70 kHz) were used for all the AFM experiments. For force reconstruction, sample rate of 1 Hz, free oscillation amplitude $\approx$ 70 nm, and trigger point of 68 nm were used. This relatively high set point enables us to avoid the bistability between the attractive and repulsive region during tip approach and yields a smooth transition between the two regimes. Amplitude A and phase $\varphi$ versus tip-sample separation distance d were recorded to employ the Sader-Jarvis-Katan formalism[12] to reconstruct the conservative forces. Since it is well-known that the tip radius R significantly affects the tip-sample interaction force, R was monitored in all experiments with critical amplitude ($A_c$) method[10] to make sure that R remains constant throughout the experiment. A minimum of 100 force profiles is reconstructed on each sample on at least 5 different locations within each sample.